\documentclass[conference, a4paper]{IEEEtran} 

\usepackage[ruled,norelsize]{algorithm2e}
\makeatletter
\newcommand{\removelatexerror}{\let\@latex@error\@gobble}
\makeatother
\ifCLASSINFOpdf
	\usepackage[pdftex]{graphicx}
	\renewcommand{\figurename}{Figure}
\else
\fi
\begin{document}

% \IEEEoverridecommandlockouts
% \IEEEpubid{\makebox[\columnwidth]{978-1-5090-4429-0/17/\$31.00 \copyright 2017 IEEE \hfill} \hspace{\columnsep}\makebox[\columnwidth]{ }}
\IEEEoverridecommandlockouts
\IEEEpubid{\begin{minipage}{\textwidth}\ \\[12pt] %\centering
\\\\\\
  Personal use of this preprint copy is permitted. Republication, redistribution\\
  and other uses require the permission of IEEE.\\
  This paper was published within the proceedings of the 14th International\\ Conference
  on Networking, Sensing, and Control (ICNSC) 2017, Calabria \\
  DOI: 10.1109/ICNSC.2017.8000113, \copyright 2017 IEEE.\\
\end{minipage}} 
\title{SensX: About Sensing and Assessment of Complex Human Motion}

\author{\IEEEauthorblockN{Andr\'{e} Ebert, Marie Kiermeier, Chadly Marouane, and Claudia Linnhoff-Popien}
\IEEEauthorblockA{Mobile and Distributed Systems Group\\
Institute for Computer Science\\
Ludwig-Maximilians-University, Munich\\
Oettingenstr. 67, 80538 Munich, Germany\\
Email: \textit{andre.ebert@ifi.lmu.de, marie.kiermeier@ifi.lmu.de, marouane@virality.de, linnhoff@ifi.lmu.de}}}


%Telephone: (800) 555--1212, Fax: (888) 555--1212}


% \IEEEpubid{0000--0000/00\$00.00~\copyright~2012 IEEE}

\maketitle

% \fi
\IEEEpeerreviewmaketitle

\begin{abstract}
The great success of wearables and smartphone apps for provision of extensive physical workout instructions boosts a whole industry dealing with consumer oriented sensors and sports equipment. 
But with these opportunities there are also new challenges emerging. 
The unregulated distribution of instructions about ambitious exercises enables unexperienced users to undertake demanding workouts without professional supervision which may lead to suboptimal training success or even serious injuries. 
We believe, that automated supervision and realtime feedback during a workout may help to solve these issues.
\par
Therefore we introduce four fundamental steps for complex human motion assessment and present SensX, a sensor-based architecture for monitoring, recording, and analyzing complex and multi-dimensional motion chains. We provide the results of our preliminary study encompassing 8 different body weight exercises, 20 participants, and more than 9,220 recorded exercise repetitions. 
Furthermore, insights into SensX’s classification capabilities and the impact of specific sensor configurations onto the analysis process are given.
\end{abstract}


%% Notes of 

\section{Introduction}\label{sec:introduction}

Today it is widely admitted that exercising regularly betters physical health and psychological well-being. 
Moreover, the risk for obesity and sufferings such as chronic diseases, hypertension, diabetes, or the Alzheimer’s disease is reduced and physically active people experience an improved quality of live as well as an increased emotional and cognitive feeling of well-being \cite{radak2010exercise}. 
Body weight exercises are performed by only using an athlete’s own body weight without artificial support \cite{lauren2011bodyweight}. 
Originally, they were carried out within school sports or special forces training. 
Now they are often seen in combination with endurance and free weight training, e.g., within Crossfit or obstacle races. 
\par
The popularity of exercising among athletes and the ubiquitous presence of mobile devices are reasons for the huge number of existing workout apps (Freeletics, etc.). 
These apps feature millions of downloads and offer customized workout planning as well as detailed instructions for challenging exercises. 
But the distribution of such information without professional supervision emerges new problems like suboptimal training success or even serious injuries (e.g., at the lower back or the human spine). 
Reasons are constantly wrong conduction of specific exercises, the absence of a warming up, or false positioning of individual extremities \cite{jones1993intrinsic}. 
Moreover, the aforementioned applications provide only a sparse possibility of exercise monitoring and no possibilities for qualitative evaluation concerning the execution of specific exercises. 
E.g., when detecting defective positions which are potentially dangerous, the user needs to be warned and instructed about which detail of the exercise was performed wrong. 
Thereby, injuries due to unsupervised training sessions could be reduced drastically and inefficient or suboptimal workouts could be avoided. 
Available sensor systems on the consumer market are already enabling simple activity recognition and tracking, such as step counting with wearables or tracking of running via GPS. 
But none these systems is capable of providing support during the conduction of complex activities and multidimensional human motion chains.
\par
We address this subject by presenting SensX, a sensorbased architecture for monitoring, recording, and analyzing complex motion chains. 
Therefore, we give an insight into related work within Section II. 
%%% TODO wording fundamental steps
In Section III, we present a generic paradigm for analyzing and assessing recurrent human motion consisting of four fundamental steps. 
These are used as a basis for the development of the SensX architecture. 
Afterwards, we explain the system’s actual state of implementation in Section IV, followed by the presentation of our preliminary study featuring 20 athletes and the system’s capabilities of recognizing specific exercises with different sensor configurations in Section V. 
Section VI deals with ongoing work and still unsolved issues.

%%%%%%%%%%%%%%%%%%%%%%%%%%%%%%%%%%%%%

\section{Analysis of human and artificial motion}\label{sec:relatedwork}
Dernbach et al. present a smartphone driven approach for recognizing human activities, arranged in two groups \cite{dernbach2012simple}. 
So called simple activities (e.g., walking) are classified correctly with a rate of over 90\%. 
So called complex activities (e.g., watering flowers) were classified correctly with a rate of
35\%-50\%. 
Strohrmann et al. examine running data of 21 participants tracked with ETHOS units \cite{strohrmann2011out}. 
In order to assess performance levels and to assist training as well as to detect fatigue, they extracted foot contact duration, foot strike type and other kinematic parameters.
\par
Chang et al. tracked 9 different weight lifting exercises by using two accelerometers, one in a glove and one worn as a belt clip \cite{chang2007tracking}. 
All output data was analyzed by using the Naive Bayes Classification (NBC) and Hidden Markov Models
(HMM), both with a success of more than 90\%. 
Still, complex movement chains which encompass individual movements of specific extremities were not examined. 
Ding et al. present another platform for free-weight exercise monitoring which is called FEMO. 
It examines strength, intensity, smoothness and continuity of RFID signals \cite{ding2015femo}. 
They analyzed 10 different weight-lifting exercises with 15 participants and achieved a successful detection rate of about 85\% with a subsequent correct classification rate of 90.4\%. 
Another system for finding, recognizing, and counting repetitive exercises is Recofit, provided by Morris et al. \cite{morris2014recofit}. 
It gathers acceleration as well as rotation data of activities with different complexity.
Subsequently, these are recognized and counted by conducting a multiclass Support Vector Machine (SVM). 
For evaluation 13 exercises of varying complexity and organized in 4 different classes were examined. 
The authors were able to recognize 95\% of the workout periods and out of these they recognized
exercises with a success rate of 96\% (within a group of 7 exercises). 
Still, Recofit is not able to track individual extremities or to provide information concerning their position.
\par
Ladha et al. introduce ClimbAX, a sensor platform for skill assessment of rock climbing which is evaluating parameters such as power, control, stability, and speed \cite{ladha2013climbax}. 
Therefore, they use a tri-axial accelerometer worn on the wrist. 
After recording, the data is down-sampled to 30Hz and split into segments capturing one move between two rest periods, respectively.
The system’s performance results for 53 climbers were compared with official results of a climbing competition – correlations could be noticed. 
ClimBSN is a system for translating accelerometer data into climbing specific measures like fluidity, strength, and endurance \cite{pansiot2008climbsn}. 
The authors use a miniaturized ear-worn 3D accelerometer as a sensor and visualize their measured and translated parameters within climbing style graphs. 
In order to isolate the most discriminative features, they use Principal Component Analysis (PCA).
\par
One of the few works which tries to achieve a qualitative analysis of weight-lifting exercises is presented by Velloso et al. \cite{velloso2013qualitative}. 
Their system consists of three sensor devices (glove, belt, and wrist-band) which encompass an accelerometer, a gyroscope and a magnetometer. 
Subsequently, they recorded 5 wrongly executed exercises and tried to classify them afterwards. 
They succeeded with a rate of 78\%. But due to the need of recording mistakes individually, the system is not very scalable and the achieved sampling rate is comparably low.
GymSkill is a system for analysis and qualitative evaluation of exercises conducted on a balance board \cite{kranz2013mobile}. 
Therefore, the authors tracked accelerometer and magnetometer data and tried to rate quality on basis of features like movement smoothness and continuity. 
For evaluation they conducted a study with 6 participants and 1200 exercises; the results ranged
from average to good based on the specific exercise setup. 
\par
Other works deal with motion analysis on basis of templating approaches. 
Ebert et al. introduced a system for segmented collision detection for vehicles by using templates and Dynamic Time Warping (DTW) \cite{ebert2016segmented}. 
Incoming inertial signals are processed in realtime and the direction as well as the impact segment of hits onto a vehicle’s surface are recognized with a success rate of more than 94\%.
\par
In summary, there are different existing approaches for analyzing human or artificial motion on basis of inertial sensor data. 
Some also aim to provide a feature based, qualitative assessment, e.g., in terms of smoothness or continuity of motions. 
Still, a real assessment of complex motions in terms of exercise safety, the detection of defective positions of extremities, prevention of injuries, as well as the optimization of training results due to automated feedback to the athletes, is missing. 
Reasons for that may be the absence of additional information concerning the position of ankles or extremities due to a lack of adequate sensors for specific extremity tracking and capable of live sensing. 
Moreover, none of the approaches mentioned above encompasses an active computing unit within the system, which makes the provision of realtime feedback unreachable.

\section{Architectural concept}\label{sec:architecturalconcept}
\begin{figure}[!t]
\centering
\includegraphics[width=3.4in]{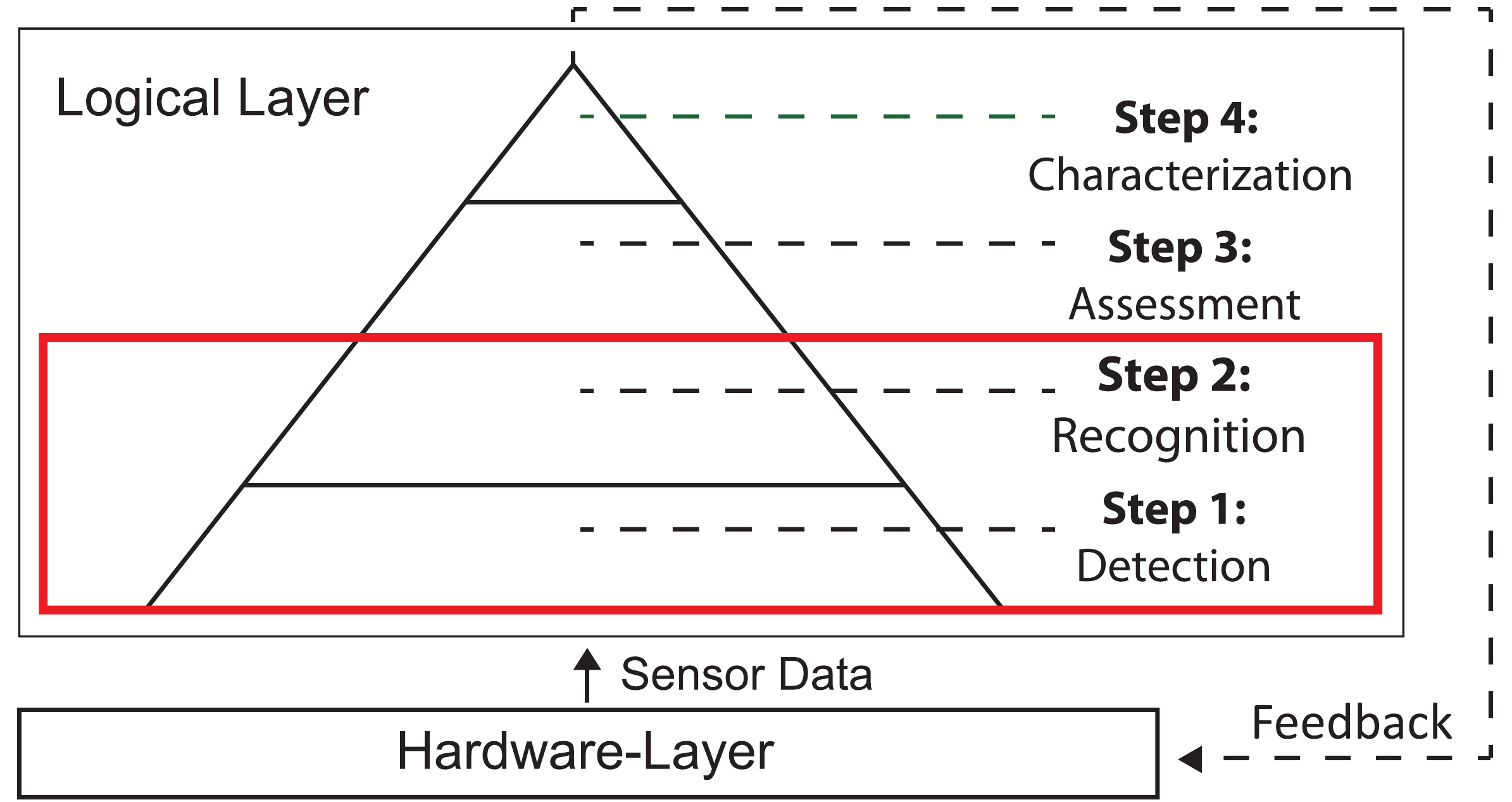}
\caption{Architectural concept with the logical layer consisting of four fundamental steps for human motion assessment and the physical layer comprising four external sensors and the central processing unit (CPU).}
\label{fig:architecturalconcept}
\end{figure}

Within this section we present the architecture of SensX, which can be split into a logical layer consisting of four fundamental steps for human motion assessment and the requirements to the system’s physical layer. 
Figure \ref{fig:architecturalconcept} depicts the overall architecture and arranges the theoretical components in context with the system’s physical properties. 
The physical layer encompasses all hardware related issues of SenseX (see Section \ref{subsec:physicallayer}). 
It tracks and distributes acceleration and rotation data which is later on used for analysis by the logical layer. 
After a successful data processing, the logical unit's results may be thrown back to the physical layer for providing feedsback to the user.

\subsection{Four fundamental steps for assessment of recurrent human motion}\label{subsec:fundamentalsteps}
\begin{figure}[!t]
\centering
\includegraphics[width=3.4in]{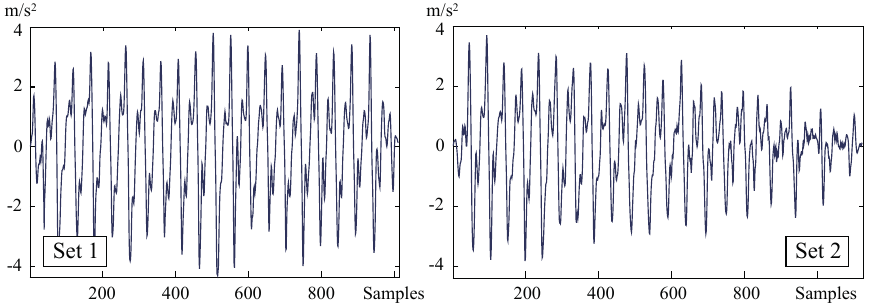}
\caption{Up and down acceleration of an athlete's chest for two sequentially conducted sets of 20 pushups each and 30s of break in between.}
\label{fig:situps}
\end{figure}
Derived from our requirements for individual human motion analysis as well as from information out of the related work mentioned above, we identified four fundamental steps (see Figure \ref{fig:architecturalconcept}, which are necessary for automatic analysis and assessment of complex chains of recurrent human motion.
\begin{enumerate}
\item \textbf{Step 1} covers the \textit{detection} of a purposeful activity out of a continuous stream of signals as well as its segmentation and preprocessing. Skipping of noise and unsubstantial sequences is simplifying later analysis.
\item \textbf{Step 2} deals with the automated \textit{recognition} of a detected activity by examining its similarity or diversity in comparison to already known classes of activities.
\item \textbf{Step 3} encompasses the \textit{assessment} process which is used to make a statement concerning qualitative predictions towards a recognized motion figure,e.g., by checking for deviations from idealized or known motion patterns or feature sets.
\item \textbf{Step 4} targets the \textit{characterization} and identification of reasons, e.g., specific anomalies, malpositions, or movements, which are leading to a potentially good or bad qualitative assessment.
\end{enumerate}
Steps 1 and step 2 as well as a possible implementation of both seems to be much more generalizable concerning their applicability for detecting and recognizing a variety of different activities. We conjecture, that the higher a step's layer is arranged within the proposed architecture, the more customization is needed to adapt it to a specific task (e.g., Step 4 characterizing a specific anomaly within an individual set of body weight exercises). Figure \ref{fig:situps} shows the acceleration of an athletes chest along the x-axis within 2 sets of 20 pushups each. It illustrates the inevitable need for customization on higher levels: a corresponding acceleration for another exercise will not contain the same qualitative information and therefore needs to be treated individually. Moreover, it demonstrates the athlete's exhaustion: the repetitions in set 1 are comparably equal – a decreasing acceleration, a longer periodic time, and striking up and down movements indicating a shivering of the athlete, are symptomatic for set 2. The following sections focus on SensX's implementation concerning step 1 and step 2, which are highlighted in Figure \ref{fig:architecturalconcept}. Step 3 and step 4 are subject of ongoing research and will be discussed in detail within future work.

\subsection{Requirements to the physical layer}\label{subsec:physicallayer}
In order to provide sufficient information for the conversion of the single steps named above we identified specific requirements which have to be met by SensX: 1) In order to identify the individual positioning of extremities and the rest of the body as well as to detect possibly defective positions, \textit{individual tracking} of extremities is needed. 
This indicates the need for a minimum of 4 sensors for the extremities and a minimum of one for the torso to enable a selective and differentiated movement analysis. 
2) As stated in \ref{sec:relatedwork}, a \textit{minimum data rate} as well as \textit{accelerometer} and \textit{gyroscope data} is needed for analysis. 
3) Only \textit{conventional hardware} is about to be used to design a system which may be affordable to ordinary athletes, disregarding of social background and financial options. 
4) An integrated \textit{central processing unit} (CPU) including computation capabilities is needed to enable realtime analysis during workouts. 
5) SensX must be \textit{physically robust} against hits and weather conditions as well as \textit{comfortable} to wear for the athletes, in order to avoid disturbances during a workout.

\section{Implementation}\label{sec:implementation}
Within these section we describe the system's current state of implementation as well as the used technologies and underlying hardware. 

%Hardware-side of the implementation
\subsection{Distributed sensor system}
As stated before, SensX needs to be able to track an athlete's extremities individually; moreover, a CPU with processing capabilities needs to be included within the system to enable immediate feedback during a workout.

\subsubsection{Central processing unit}\label{subsubsec:centralcomp}
Central processing unit: As a CPU we experimented with two different Android smartphones: A HTC One (M7) with Android 5.0 Lollipop and a Google Nexus 5x by LG running Android 6.0.1 Marshmallow. 
Our motion-tracking app enables the labeling of 8 different exercises and controls the compliance of mandatory breaks in between individual workout sets automatically. 
The tracked workout sets are currently stored row-based within individual text files, organized by the specific sensor used for tracking and by providing a timestamp for each data fix. 
The smartphones are also used as a motion sensor (accelerometer and gyroscope) and provide data-rates of roughly 150Hz (HTC) / 100Hz (LG). 
Due to different Bluetooth Low Energy (BLE) implementations (deviating from official specifications) the achieved sampling rates when connecting 4 external sensors were significantly different. 
In context of the LG we achieved a stable sampling rate for each sensor board of 50Hz (only 3-axial accelerometer, no gyroscope), the HTC performed with a 40Hz sampling rate for both, four 3-axial accelerometers and four 3-axial gyroscopes.

\subsubsection{External devices}\label{subsubsec:externalsensors}
Small, multi-functional, highly-usable, and robust sensors at affordable prices and suitable for analyzing human motion under harsh conditions (e.g., humidity, rough motions, etc.) are only available since a short time.
SensX uses 4 MBIENTLAB Meta Wear CPRO sensors, containing an accelerometer, a gyroscope, a magnetometer, a barometer and an ambient light sensor. 
In context of this work, we recorded 30 different signal vectors: x-, y-, and z-axis of acceleration and rotation for four external sensors plus the smartphone's sensor set. 
The theoretical sampling rate for the Meta Wear boards is more than 100Hz per sensor, but due to BLE related restrictions, the achieved data rate is lower (see Section \ref{subsubsec:centralcomp}).

\subsubsection{Data transfer}
Data may be stored on a small 256kB storage or transferred directly to the processing unit via BLE.
Due to the required realtime feedback and the necessity for processing bigger amounts of data while tracking an exercise we use the BLE connection for connecting all four sensor boards.
Hardware related differences for the achieved sampling rates are described in Section \ref{subsubsec:centralcomp}.

\subsubsection{Sensor attachment}\label{subsec:sensattach}
\begin{figure}[!t]
\centering
\includegraphics[width=3.4in]{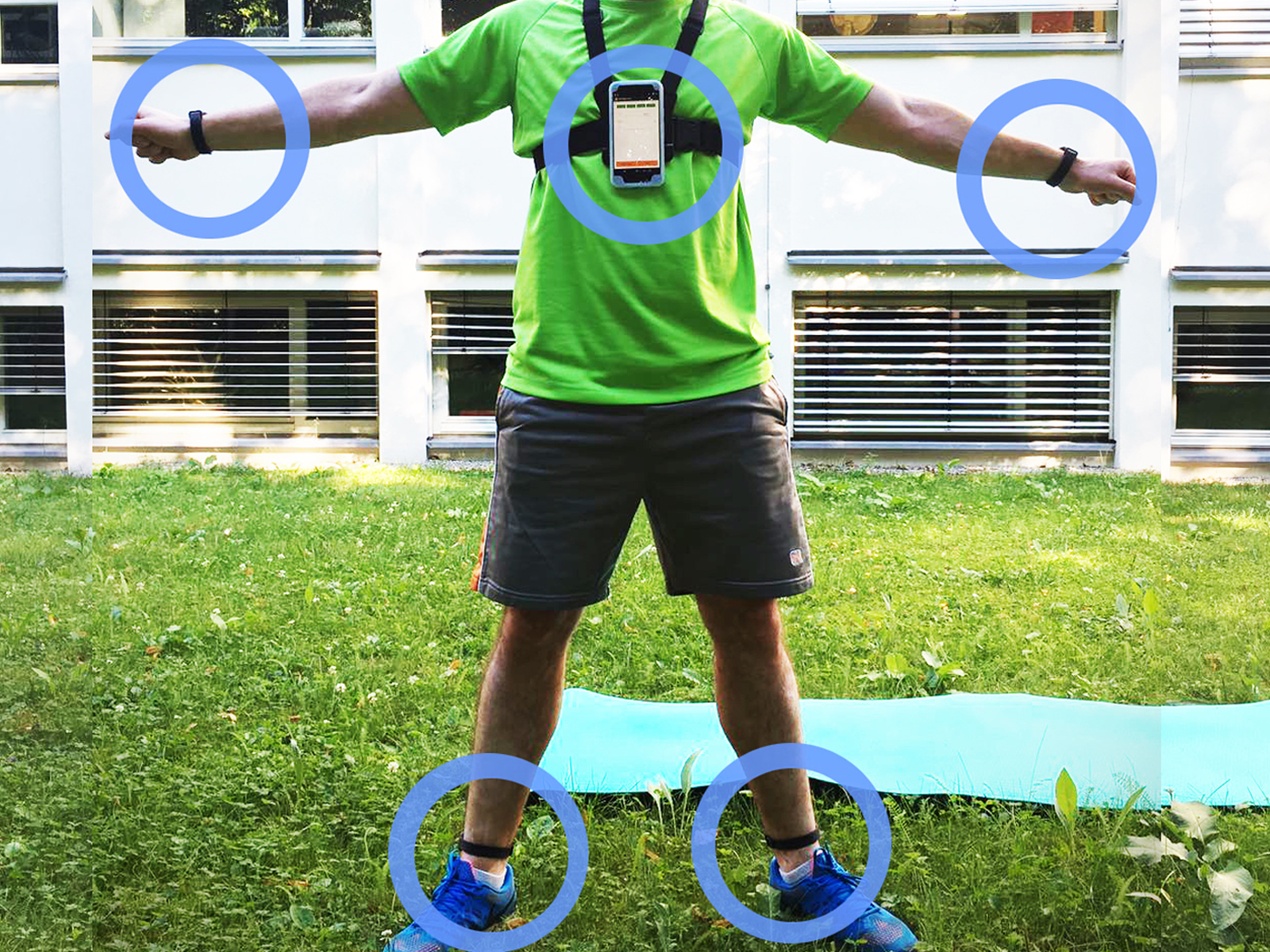}
\caption{Attachment of the SensX sensor system on the athlete's body.}
\label{fig:attachment}
\end{figure}

The processing unit is mounted with a GoPro harness on the athletes' chest, the display pointing forward. 
The external sensor devices are packed within a plastic case fastened with rubber bands on each individual extremity (see Figure \ref{fig:attachment}). 
Each external sensor has a mark concerning its orientation in order to ensure a correct attachment at all times.

%% Software-sided implementation
\subsection{Logical Layer}\label{subsec:loglayer}
In order to detect and extract single exercises into frames, an advanced version of the peak-detection algorithm presented in \cite{ebert2016segmented} is used. 
In contrast to the homogeneous profile of directional impacts on a vehicle's surface, signals of body weight exercises may vary significantly in their length, continuity, wave shape, and complexity. 
To respect that fact, we extended the proposed algorithm and use the signal with the greatest steady dynamics and identify the first occurrence of a periodic exercise as well as its initial frame-length from its autocorrelation.
Subsequently, every occurrence is checked for its zero-crossings and the frame size is enlarged if necessary.
For smoothing, a Butterworth lowpass is used again. 
A new problem occurring by using this approach is the techincally conditioned capability for realtime processing: in order to process the auto-correlated signal, a sequence of exercises is needed. 
This issue could be solved in future by using initial frame lengths which are determined statistically and become adjusted over time.
To realize the recognition of different exercises we experimented with the Naive Bayes Classification (NBC) as well as with the J48 algorithm, an Open-Source implementation of the C4.5 decision tree algorithm which is capable of multidecisioning (in contrast to e.g., binary CART capabilities), both in combination with the WEKA machine learning framework \cite{drazin2012decision}. 
Again, the system still lacks of realtime capabilities and the analysis is currently conducted offline by using a separate machine. 
For the following evaluation in Section \ref{sec:evaluation} we only used NBC due to a significantly faster runtime as well as constantly better results for our use-case.

\section{Evaluation}\label{sec:evaluation}
\begin{figure}[!t]
\centering
\includegraphics[width=3.4in]{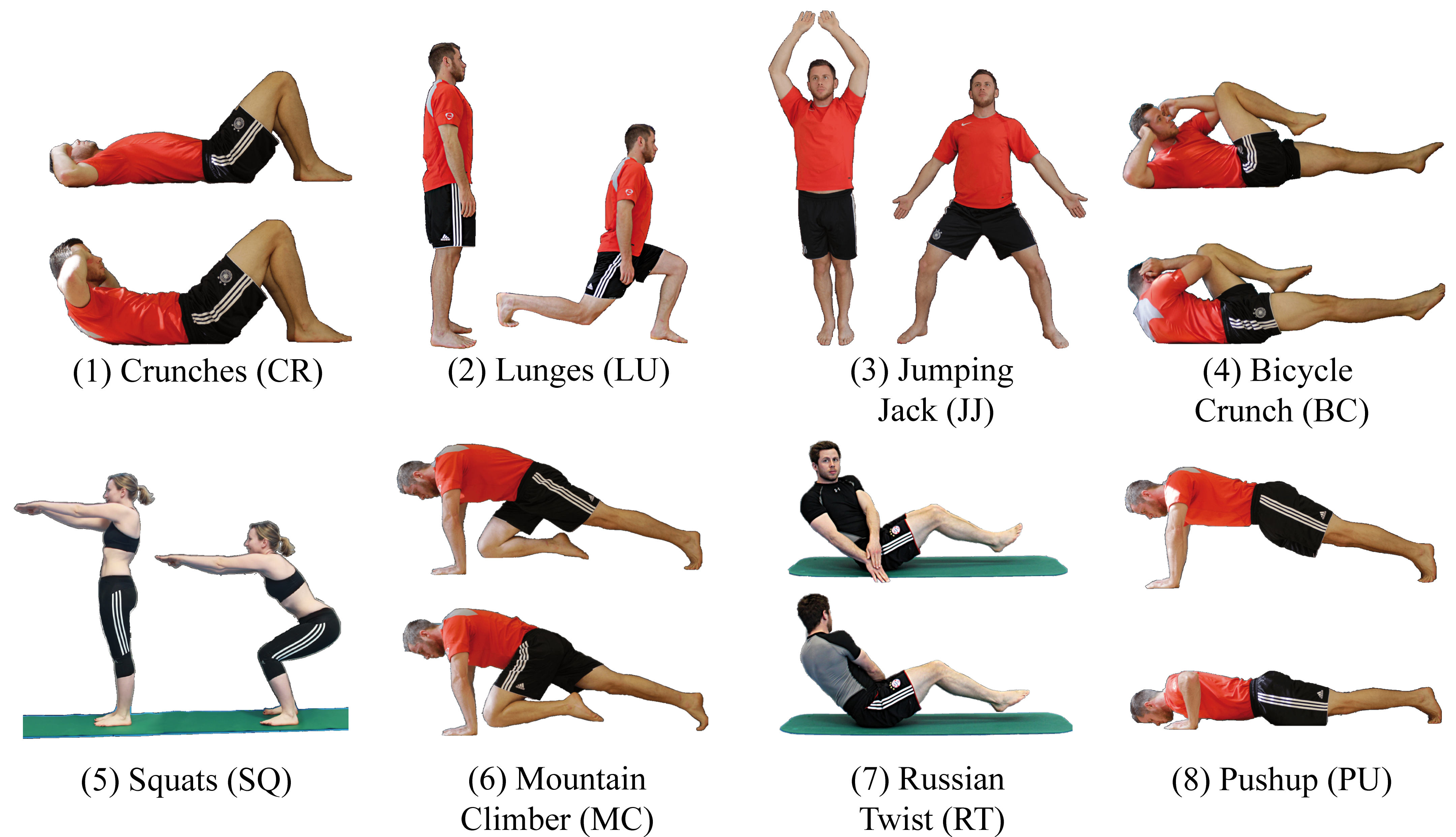}
\caption{All exercises which were executed within the study ordered by their conduction and including their abbreviations.}
\label{fig:exercises}
\end{figure}
Subsequently, we present the results of a preliminary study concerning the system's capabilities of recognizing different exercises as well as the performance of different sensor configurations. 
As a CPU we only used the Nexus 5x in combination with four external sensors.
%% TODO note only acceleration because of data rate

\subsection{Study design}\label{subsec:studydesign}
Within the study's context we tracked the execution of 8 different body weight exercises (Crunch (CR), Lunge (LU), Jumping Jack (JJ), Bicycle Crunch (BC), Squat (SQ), Mountain Climber (MC), Russian Twist (RT), and Pushup (PU), see Figure \ref{fig:exercises}) carried out by 20 amateur athletes with an average age of 24 years, whereby 20\% were of female and 80\% of male sex. 
The average Body Mass Index was $22,32kg/m^2$ and the average height was $178,2m$. To ensure the completion of the whole workout by as much athletes as possible and to avoid muscular congestion, the exercises' order was designed to stress different groups of muscles, successively. 
Each exercise was instructed with a video, then the athlete had to conduct 3 sets with 20 repetitions; between the individual sets there was an obligatory break of $30s$. 
All in all we recorded more than 9,220 repetitions (approximately 1,152 repetitions per exercise), not all athletes managed to complete the whole workout.
Besides tracking the athlete's acceleration data all workouts were also captured on video to enable additional analysis later on. 
Furthermore, we surveyed the personal information about each athlete and the subjectively felt individual performance.
Anomalies occurring during the workout were also noted.

\subsection{Analysis and data splitting}\label{subsec:datasplitting}
The configuration tests described within Section V-C were conducted with a training vs. test data ratio of 80:20 as proposed by Ng \cite{ng2016datasplitting}. 
For the evaluation in Section \ref{subsec:generalclassification} we used the same ratio for the training/test datasets, for k-fold cross validation (k = 4) we used the training dataset.

\subsection{Sensor configurations}\label{subsec:configurations}
\begin{figure}[!t]
\centering
\includegraphics[width=3.5in]{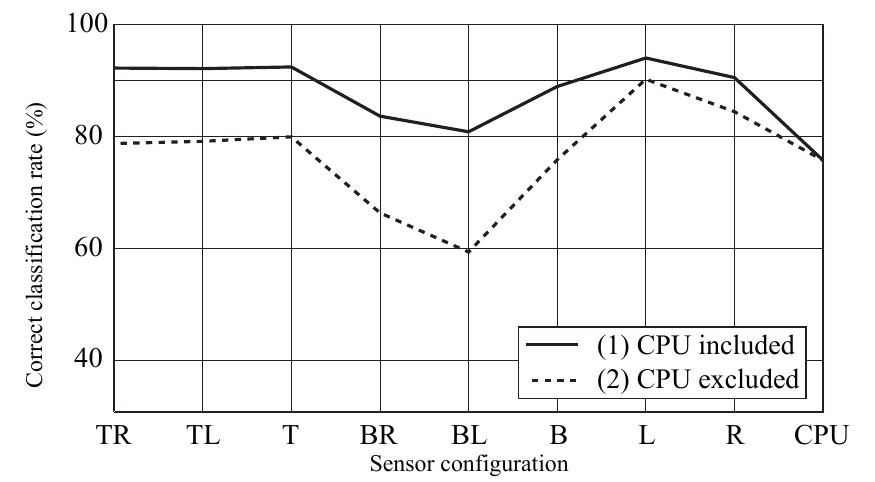}
\caption{Recognition performance for different sensor configuration with (1) and without (2) inclusion of sensor data provided by the CPU.}
\label{fig:recpersens}
\end{figure}

\begin{table}[htb]
\scriptsize
\centering
\begin{tabular}{|r|c|c|c|c|}
\hline
\textbf{Abbrev.} & \textbf{wrist left} & \textbf{wrist right} & \textbf{foot left} & \textbf{foot right}\\ \hline
ALL & $\surd$ & $\surd$ & $\surd$ & $\surd$\\ 
top right (TR) & $\times$ & $\surd$ & $\times$ & $\times$\\ 
top right (TL) & $\times$ & $\surd$ & $\times$ & $\times$\\ 
top (T) & $\surd$ & $\times$ & $\times$ & $\times$\\
bottom right (BR) & $\times$ & $\times$ & $\times$ & $\surd$\\ 
bottom left (BL) & $\times$ & $\times$ & $\surd$ & $\times$\\
bottom (B) & $\times$ & $\times$ & $\surd$ & $\surd$\\ 
left (L) & $\surd$ & $\times$ & $\surd$ & $\times$\\ 
right (R) & $\times$ & $\surd$ & $\times$ & $\surd$\\ \hline
\end{tabular}
\vspace{-2mm}
\\\bigskip
\begin{flushleft}
Table \ref{tab:sensorconfigs}1. Different sensor configurations and their abbreviations.
\end{flushleft}
\label{tab:sensorconfigs}
\end{table}
\vspace{-2mm}
To examine different sensor configurations, we created 9 different sensor groups which are depicted in Table \ref{tab:sensorconfigs}.
These are encompassing different regions of the human body, such as top (both arms), bottom (both legs), top right (right arm), or right (right arm + right leg). Subsequently we determined the successful recognition rate for each individual sensor configuration. 
Figure \ref{fig:recpersens} presents the classification results for using sensor data of the specific sensor groups and including
(1) as well as excluding (2) additional acceleration and rotation information provided by the CPU. 
The performance for tracking only a single leg (BR, BL) is the worst with success rates lower than 70\%.
In contrast to that, the tracking of only one arm (TL, TR), both arms (T), or both legs (B) provides better results, but still retrieves a successful recognition of activities in less than 80\%. 
The tracking of the whole left (L) or the whole right side (R) of an athlete performs significantly better with success rates in between 80\% and more than 90\%, respectively. 
When only the CPU without any external sensor data is used, a rate of 75.5\% can be achieved. 
(2) shows the results for all configurations in combination with the CPU. 
All of them reach a correct classification rate of more than 80\%, the groups TR, TL, T, L, and R perform with a success rate of even more than 90\%. 
The results show, that information provided by the bottom extremities contains less valuable information for exercise classification than from the top. 
Moreover, the acceleration data of only two external sensors in combination with a chest sensor is enough to reach recognition rates of more than 90\%, although there is no additional information about the position of all individual extremities.
\par
The overall correct classification rate when using all sensors together was about 95.2\%. 
The proposed full sensor configuration setup with four external sensors performs perceptible better compared to reduced configurations.

\subsection{General exercise recognition}\label{subsec:generalclassification}
After presenting performance differences in between different sensor groups within Section \ref{subsec:configurations}, we now examine the differences within results for specific exercises. 
Therefore, we conducted a new NBC featuring two test cycles: one cycle conducts a k-fold cross validation a second cycle uses training and test data (see Section \ref{subsec:datasplitting}).
Table \ref{tab:exercisespecificres} presents the average rate of successful exercise recognitions for both cycles. 

\begin{table}[htb]
\scriptsize
\centering
\setlength{\tabcolsep}{0.5em} % for the horizontal padding
{\renewcommand{\arraystretch}{1.2}% for the vertical padding
\begin{tabular}{|l|l|l|l|l|l|l|l|l|l|}
   \hline
 & \textbf{CR} & \textbf{LU} & \textbf{JJ} & \textbf{BC} & \textbf{SQ} & \textbf{MC} & \textbf{RT} & \textbf{PU} & \textbf{Avg.} \\ 
	\hline
CV & 96.9\% & 97.5\% & 97.5\% & 97.3\% & 96.7\% & 97.4\% & 97.9\% & 75.9\% & 94.6\% \\ 
TD & 98.0\% & 96.3\% & 98.4\% & 93.7\% & 97.5\% & 96.7\% & 93.1\% & 82.0\% & 94.5\% \\
\hline
\end{tabular}
}
\vspace{-2mm}
\\\bigskip
\begin{flushleft}
Table \ref{tab:exercisespecificres}. Successful classification rates (in \%) for specific exercises when using cross validation (CV) and testdata (TD).
\end{flushleft}
\label{tab:exercisespecificres}
\end{table}
\vspace{-2mm}

\begin{figure}[!t]
\centering
\includegraphics[width=3.4in]{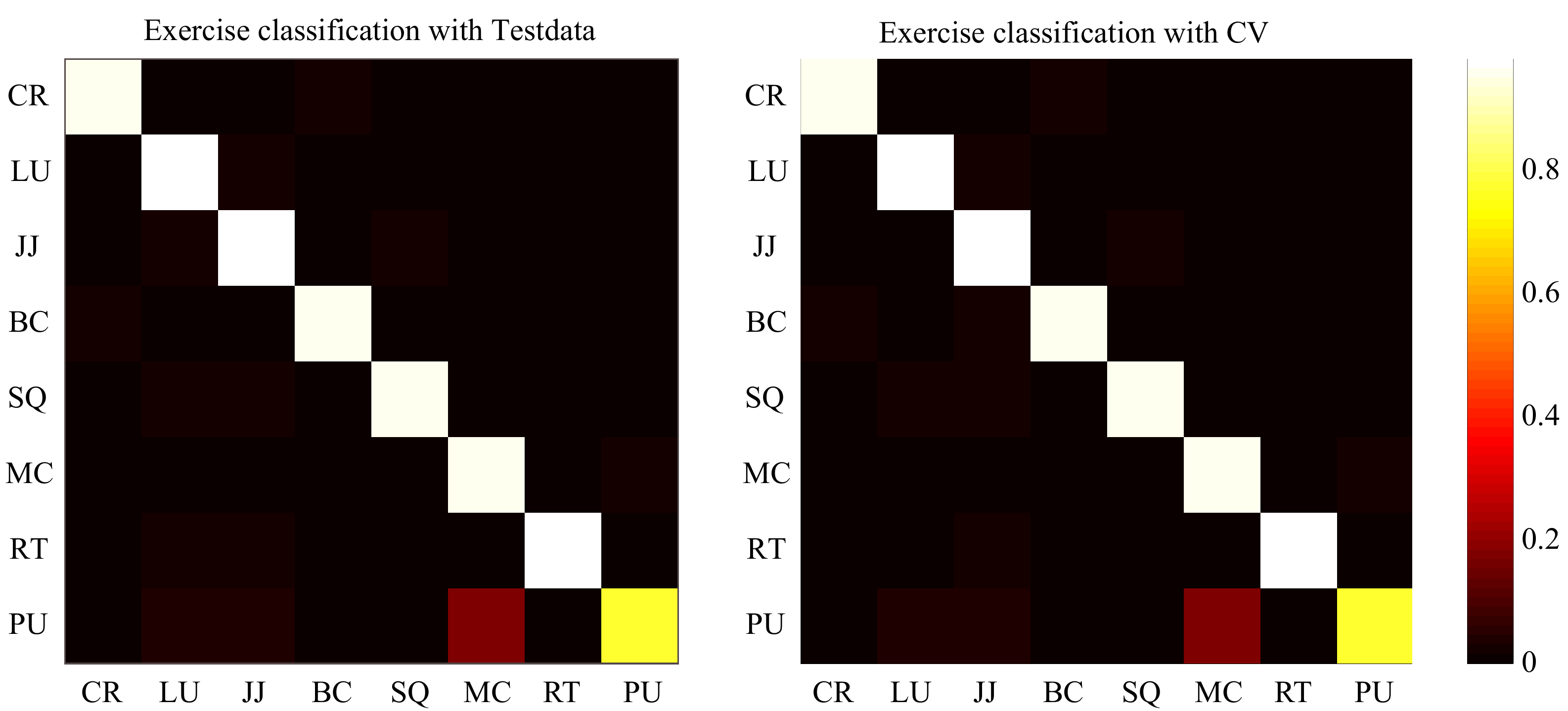}
\caption{Confusion matrices for exercise recognition with cross validation and for using training/test set.}
\label{fig:exerciseconfusion}
\end{figure}

The results for both cycles show a high amount of similarity, which is also indicated by the confusion matrices depicted in Figure \ref{fig:exerciseconfusion}. 
It is significant that for all exercises except pushups within both cycles the successful recognition rate is far above 90\% – pushups are only recognized with a rate of 75.9\% and 82.0\%.
All in all, pushups were mostly mistaken with the mountain climber exercise. We believe, that there are 3 reasons for that: 1) the starting position as well as the general position for both exercises is very similar (see Figure 4), 2) both exercises were extremely exhausting compared to others so that motions were not always conducted as distinct as for other exercises, and 3) because of the pushups’ scheduling at the end of the workout all athletes were exhausted when coming to that challenging exercise. 
As a consequence of that, only a few managed to conduct 20 clean pushups for all 3 sets. 
This resulted in fewer and more noisy pushup datasets. 
Furthermore, the presence of rotation information could have given essential information for recognizing pushups more distinct from the mountain climber.
The average results are more than 94\% when using training data as well as when using cross validation. If we only validate data of 7 exercises of even data quality and quantity (without pushups) the successful classification rate is 96.2\% for using data and 97.3\% cross validation.

\subsection{Miscellaneous characteristics and findings}
SensX meets all requirements stated within Section \ref{subsec:physicallayer}.
Due to its small and flexible external sensors it is capable of tracking the movements of all individual extremities with a joint sampling rate of more than 400Hz, split in 30 different motion signals. 
Concerning the system's attachment, all athletes stated that they felt comfortable and not hindered by the
sensors during working out.

\section{Conclusion and future work}\label{sec:conclusion}
In context of this paper we presented a paradigm for analyzing and assessing recurrent, complex human motion
consisting of four fundamental steps.
Subsequently, we introduced SensX, an architecture for tracking, analyzing and assessing human motion as well as its current state of implementation and the results of our preliminary study for activity recognition.
The system consists of one central processing unit and four external sensors that track acceleration
as well as rotation data. 
Moreover, it is able to track all four human extremities individually. 
Furthermore, we presented a study with 20 athletes conducting 8 different body weight exercises and more than 9,220 individual repetitions. 
The successful exercise classification within our evaluation proofs that SensX already provides convincing results when only using acceleration information. 
Besides that, we also examined the performance of different sensor configurations and learned that tracking only one side of the human body plus the CPU data is enough information for getting results which are comparable to using the complete sensor set (but only for recognition).
While step 1 and 2 of the proposed logical layer are implemented by SensX, step 3 and 4 which deal with qualitative exercise assessment are still unsolved and part of our ongoing work. 
In that context, we conducted a new study with nearly 30 participants using the HTC configuration (see \ref{subsubsec:externalsensors}) and tracked acceleration as well as rotation data. 
Again, all workouts were video taped. 
As shown in \cite{ebert2016segmented}, rotation data contains crucial information about complex movements and we believe that this additional information together with the one of individual limb movement will enable us to create valid quality assessment processes for human motion.
\par
Another unsolved problem is the porting of our analysis approach towards a live capable system, which may provide feedback to the user in realtime. 
By using small, flexible and comfortable-to-wear external sensors, the system seems to be fit for realtime in hardware terms, but the analysis runtime, the identification of adequate and economical algorithms for quality assessment as well as the fact that exercises must be processed sequentially from a continuous data stream rises up new challenges. 
Here, the usage of distance measurements or more simplified alternatives to machine learning approaches may be the solution.
Based on the learnings of this work, we believe that efficient recognition and assessment of body weight exercises as well as the provision of specific feedback to the user is reachable.

\bibliographystyle{IEEEtran}
\bibliography{refs}



%\end{thebibliography}

\end{document}